\documentclass[preprint2]{aastex}
\begin{document}

\title{Pregalactic Black Hole Formation with an Atomic Hydrogen Equation of State}

\author{Marco Spaans\altaffilmark{1} and Joseph Silk\altaffilmark{2}}

\affil{$^1$Kapteyn Astronomical Institute, P.O. Box 800, 9700 AV
Groningen, The Netherlands; spaans@astro.rug.nl}
\affil{$^2$Astrophysics, Oxford University, 1 Keble Road, Oxford OX1
3RH, United Kingdom; silk@astro.ox.ac.uk}

\begin{abstract}

The polytropic equation of state of an atomic hydrogen gas is
examined for primordial halos with baryonic masses of $M_h\sim 10^7-10^9$
$M_\odot$. For roughly isothermal collapse around $10^4$ K, we find that
line trapping of Lyman $\alpha$ (HI and HeII) photons causes the polytropic
exponent to stiffen to values significantly above unity. Under the assumptions
of zero H$_2$ abundance and very modest pollution by metals ($<10^{-4}$ Solar),
fragmentation is likely to be inhibited for such an equation of state. We
argue on purely thermodynamic grounds that a single black hole of
$\sim 0.02-0.003M_h$ can form at the center of a halo for $z=10-20$
when the free-fall time is less than the time needed for a resonantly
scattered Lyman $\alpha$ photon to escape from the halo. The absence of H$_2$
follows naturally from the high, $>10^4$ K, temperatures that are attained
when Lyman $\alpha$ photons are trapped in the dense and massive halos that
we consider. An H$_2$
dissociating UV background is needed if positive feedback effects on H$_2$
formation from X-rays occur.
The black hole to baryon mass fraction is suggestively close to
what is required for these
intermediate mass black holes, of mass $M_{BH}\sim 10^4-10^6$ $M_\odot$, to
act as seeds for forming the supermassive black holes of mass 
$\sim 0.001M_{spheroid}$ found in galaxies today.

\end{abstract}


\section{Introduction}

A fundamental issue in the study of galaxy evolution is the formation
of the central (supermassive) black hole.
Accretion onto these black holes provides the energy source for active galactic
nuclei, which in turn impact the evolution of galaxies (Silk 2005).
Earlier attempts at providing seeds for galactic black holes include
dynamical friction and collision processes in dense young stellar clusters
(Portegies Zwart et al.\ 2004) and formation from low
angular momentum material in primordial disks (Koushiappas, Bullock \&
Dekel 2004).

In this work, we consider the impact of the polytropic equation of state (EOS)
of a metal-free, atomic hydrogen gas on the expected collapse of matter
inside massive halos.
The impact of a Solar metallicity polytropic EOS on the expected masses of
stars in local galaxies has been
investigated by Li, Mac Low \& Klessen (2003). The influence of molecular
hydrogen and
metal-poor environments has received detailed interest from, e.g.,
Abel, Bryan \& Norman (2002, 2000); Bromm, Coppi \& Larson (2002) for the
formation of the first stars and from
Scalo \& Biswas (2002) and Spaans \& Silk (2005) for the properties of the
polytropic equation of state.
From the work of Li et al.\ (2003) it has become clear that a polytropic EOS
of state, $P\propto\rho^\gamma$ with $\rho$ the mass density and $\gamma$
the polytropic exponent, strongly suppresses fragmentation of interstellar gas
clouds if $\gamma >1$.
This paper concentrates on the impact of Lyman $\alpha$ photon trapping on
the EOS and the interested
reader is referred to Rees \& Ostriker (1977) and Silk (1977) for some of the 
fundamental
thermodynamic and star formation considerations that come into play here.

\section{Model Description}

We assume a metal-free hydrogen gas that is cooled
by Lyman $\alpha$ emission as it collapses inside a dark matter halo and
radiates away about twice its binding energy
(Haiman, Spaans \& Quataert 2000). Note that Lyman $\alpha$ cooling is
expected to dominate over radial contraction factors of at least 15-60 as long
as the
metallicity is less than 0.1 of Solar (Haiman et al.\ 2000). The absence of
any H$_2$, that would cool the gas to below 8,000 K, is crucial in this
and we come back to this point in Section 4. We further employ
a polytropic EOS and a perfect gas law, $P\propto\rho T$, for the gas
temperature $T$, and write $\gamma$ as
$$\gamma =1+{{d{\rm log}T}\over{d{\rm log}\rho}}. \eqno(1)$$
This last step is justified (Scalo \& Biswas 2002) as long as the heating and
cooling terms in the
fluid energy equation can adjust to balance each other on a time-scale shorter
than the time-scale of the gas dynamics (i.e., local thermal equilibrium).
Below, we compute the polytropic EOS for the case that the
cooling time is shorter than the free-fall time and for the case that the
photon propagation time exceeds the dynamical time.

It should be noted that because $\gamma$
depends on the (logarithmic) derivative of the temperature with
respect to density, it implicitly depends on radiative transfer effects and
changes in chemical composition through derivatives of the heating and
cooling functions (Spaans \& Silk 2000). The Lyman $\alpha$ radiative transfer
techniques as described in Haiman \& Spaans (1999) and Dijkstra et al.\ (2006)
are used to compute the transfer of Lyman $\alpha$ photons.

We consider spherical dark matter halos that have decoupled from the Hubble
flow and are characterized by a mean density of $\rho\approx 200\rho_b(1+z)^3$
at $z=10-20$, for a baryonic number density
$\rho_b/m_H=3\times 10^{-7}$ cm$^{-3}$ today, hydrogen mass $m_H$,
total halo masses of $M_h=10^7-10^9$ $M_\odot$ and a characteristic size scale
of $L=(3M_h/4\pi\rho)^{1/3}$. This yields, over $z=10-20$, a typical mean
density and column of $n_0=0.05[(1+z)/10]^3$ cm$^{-3}$ and
$N_0=10^{22}[(1+z)/10]^2(M_h/10^9M_\odot)^{1/3}$ cm$^{-2}$,
respectively. We further assume that matter inside the halo remains at
approximately $10^4$ K during its collapse so that an isothermal density
profile, $n\propto n_0(L/r)^2$, is applicable for every radius $r$.
Therefore, the column
a Lyman $\alpha$ photon has to traverse from a radius $r$ to $L$ scales
as $\int_{r}^Lndr\sim L/r-1$ with a mass inside of $r$ of $M(r)\sim r$.

In the absence of any ionizing sources, heating is provided by gravitational
compression, $\Gamma\propto n^{1.5}$.
The velocity dispersion of the gas is thermal and equal to $\Delta V =12.9T_4$
km/s, with $T_4$ in units of $10^4$ K. The natural to thermal line width of
the Lyman $\alpha$ line is
denoted by $a$ and equal to $a=4.7\times 10^{-4}T_4^{-1/2}$.

\section{Results}

\subsection{Static Case}

With cooling provided by Lyman $\alpha$ emission only, the thermal equilibrium
of the baryonic matter in the halo approximately (Spitzer 1978; Haiman et al.\
2000) follows, for $r'$ in units of L,
$$7.3\times 10^{-19}n_e(r')n_H(r')e^{-118,400/T(r')}\epsilon (r')=$$
$$1.9n(r')GM_h/L t_{ff}^{-1},\eqno(2a)$$
with $M_h/L=M_h(r'/L)/r'=M(r')/r'$,
electron density $n_e$, atomic hydrogen density $n_H$, Lyman $\alpha$
escape fraction $\epsilon$ and free-fall time
$$t_{ff}=4.3\times 10^7/n(r')^{1/2} {\rm yr}.\eqno(2b)$$
It is assumed here that the cooling
time at the peak of the cooling curve is $t_c=3/2nkT_{vir}/n^2\Lambda$, with
Boltzmann's constant $k$, the halo's virial temperature $T_{vir}$ and
$\Lambda\approx 2\times 10^{-22}$ erg s$^{-1}$ cm$^3$. Depending on the ambient
conditions, the medium cools somewhere around the peak of the cooling curve
and $T_{vir}<10^4$ K if the mass is smaller than
$10^8[(1+z)/10]^{-1.5}$ $M_\odot$ for the virialization redshift $z$ (Haiman,
Rees \& Loeb 1997).

The escape fraction $\epsilon$ of Lyman $\alpha$ photons from a sphere
diminishes from unity when collisional de-excitation above a critical HI
column density $N_c$ becomes important (Neufeld 1990). This column $N_c$
depends on the ambient temperature and
ionization balance through the probability for collisional de-excitation
$p_0={{q_pn_p+q_en_e}\over{A_{21}}}$, with proton density $n_p$, collisional
de-excitation rate coefficients $q_p$
and $q_e$, and the Einstein $A$ coefficient $A_{21}$ connecting the $2p$ and
$2s$ states. In this, the ambient proton density is assumed to be lower than
about $10^4$ cm$^{-3}$ so that the created $2s$ hydrogen atoms undergo
two-quantum decay to the ground state.
It follows that $N_c$ ranges between $10^{21}$ and $10^{23}$
cm$^{-2}$ for $y=n_pT_4^{-0.17}$ between $10^2$ and $10^4$ cm$^{-3}$,
respectively (Dijkstra et al.\ 2006), and is much larger for much smaller
proton densities. These values for $N_c$ are a factor of
a few larger than the corresponding values for a slab (Neufeld 1990), since
resonantly scattered photons escape
more easily from a sphere than from a slab for the same surface-to-center
optical depth.

Furthermore,
following the Monte Carlo radiative transfer techniques in Dijkstra et al.\
(2006) and Haiman \& Spaans (1999), and for a HI column $N_H$, we find that
$\epsilon\approx N_cN_H^{-1.0}$, for $N_H=2N_c-100N_c$ and for spherical
clouds.
In deriving this fit to the numerical results, we have made sure that the line
profile is sampled far enough into the wings to accurately determine
$\epsilon$ and $N_c$.
When applied to a slab, rather than a sphere (see the analytical solution in
the appendix of Dijkstra et al.\ 2006), our method yields results that
agree well with those of Neufeld (1990, his Figure 18).

For the resonantly scattered Lyman $\alpha$ line, it follows,
for a line center optical depth $\tau_0$, mean line opacity $\alpha_s$ and
profile function
$\phi (x)$ in normalized frequency units $x$, that an escaping photon,
which scatters $N$ times, experiences a frequency shift $x_s\sim N^{1/2}$ and
travels a distance $N^{1/2}/(\alpha_s\phi (x_s))$ that is equal to the size of
the medium $\tau_0/\alpha_s$. Hence,
$x_s\sim\tau_0\phi (x_s)\sim (a\tau_0)^{1/3}$ since $\phi\sim a/\pi x^2$. On
average, a time $\delta t\sim {{L}\over{c}}/(\tau_0\phi (x_s))$ elapses
between the $\sim N$ scatterings. Thus, a time
$t_{ph}\sim N\delta t\sim{{L}\over{c}}(a\tau_0)^{1/3}$ is required for a
photon to escape,
where the optical depth is given by $\tau_0 =1.04\times 10^{-13}N_HT_4^{-1/2}$.
Typically, we have that $\tau_0 >10^7$. Thus, for a given density, and in the
limit that $t_{ff}>>t_{ph}$ with $N_H>N_c$, an increase in column leads to a
proportional decrease in the spherical escape probability
$1-e^{-a\tau_0}/a\tau_0 \approx 1/a\tau_0$
and the chance that a scattering hydrogen atom will not suffer collisional
de-excitation effects.

Obviously then, Equation (1) has a weak dependence of $\gamma$ on density
for modest columns due to the exponential temperature dependence and
the $N_H^{-1.0}\propto n_0^{-2/3}$ scaling of $\epsilon$ in the static case
and for fixed mass $M_h$. Under collisional ionization equilibrium, one finds
from solving Equations (2) for $T(n)$ that
$$\gamma -1\approx -{{1}\over{2log(Cn^{1/2})}}=0.006-0.007,\eqno(3)$$
for proton
densities larger than $10^{2-4}$ cm$^{-3}$, $N_H>10^{21-23}$ cm$^{-2}$ or
$r\le r_{stat}=1.0-0.01L$ for all halos over $z=10-20$, and where
$C\sim 10^{-36}$ $M_h/10^7M_\odot$ cm$^{3/2}$. Hence, as to be expected, the
stiffening of the polytropic EOS is always modest when the exponential
temperature dependence of the Lyman $\alpha$ cooling rate acts unchecked.

\subsection{Dynamic Case: HI}

For the halos considered here, $t_c<t_{ff}$ by a factor of a few. However,
if the random walk that a Lyman $\alpha$ photon performs takes a time $t_{ph}$
that is comparable to or longer than
the dynamical time on which the halo evolves, cooling is effectively shut down.
Photons can then only escape through parts of the line wings that have modest
optical depths, while the Lyman $\alpha$ emission becomes zero around line
center.

One can show that
$$\epsilon\rightarrow\epsilon\times e^{-\beta t/t_{ff}},\eqno(4)$$
with $t=t_{ph}$, as more and more photons get trapped in the line core for
times exceeding the dynamical time. The multiplier $\beta\sim 2-3$ incorporates
details of the gravitational collapse of gas shells
(e.g.\ geometry, kinematics) and does not impact our results
as long as it does not (or only weakly) depend on density.

That is, the decrease in the number of escaping/cooling Lyman $\alpha$ photons
is approximately proportional to the total number of photons somewhere in the
line multiplied by the average time a given photon spends in the medium per
unit of free-fall time: gravitational collapse scales with
$t_{ff}\propto n^{-1/2}\propto r$ and the number of already collapsing shells
a photon would have to traverse thus increases linearly in space and time,
i.e., $-d\epsilon\sim\epsilon dt/t_{ff}$.
It is implicitly assumed here that the scattering--broadened line width,
typically larger than $200$ km/s (Dijkstra et al.\ 2006), exceeds any
systematic velocity shifts, which is a good approximation for large optical
depths.
Hence, in Equation (2a) the factor $\epsilon$ is now competitive with the
temperature dependence of Lyman $\alpha$ cooling because it picks up an
exponential function of density.

Typically, one has $t_{ff}\sim 1.6\times 10^{15}/n_0^{1/2}$ s and
$t_{ph}\sim 5.0\times 10^{14}/n_0^{1/9}(M_h/10^9M_\odot)^{1/3}$ s for the
adopted halo characteristics. Note in this expression, 
the weak and negative dependence
of $t_{ph}$ on density. This is a consequence of the random walk in both
coordinate and frequency space that is performed by the Lyman $\alpha$ photon,
yielding a weak $\tau_0^{1/3}\sim n_0^{2/9}$ dependence for $t_{ph}$, while
the size of the halo scales as $n_0^{-1/3}$ for a fixed mass $M_h$.

The expression for the local thermal balance, Equation (1) above, formally
does not change, although all quantities pick up a time dependence, as long as
local thermal balance holds. This is still true for $t_{ph}\ge t_{ff}$ and
$T\sim 10^4$ K, given that the time needed to thermalize through collisons
scales as $1/n$ and the free-fall (heating) time as
$1/n^{1/2}$. Similar considerations apply to the ionization balance of
hydrogen, but the presence of shocks would require a more careful treatment.
The velocity gradients that exist maximally have a magnitude of
$$\delta v\sim r/t_{ff}(r)\sim 10^2\ {\rm km/s},\eqno(5)$$
smaller than the scattering broadened line width, and are independent of $r$
if an isothermal density distribution pertains.

One can determine $\gamma$ straightforwardly for
$t\sim t_{ff}$ and a fixed mass $M_h$. One finds that
$$\gamma -1\approx -{{{{1}\over{2}}+{{7}\over{18}}Bn^{7/18}}\over{log(Cn^{1/2})+Bn^{7/18}}},\eqno(6)$$
where it should be noted that $t_{ph}/t_{ff}\propto n^{7/18}$ and that
$B\approx 0.5-0.1$ cm$^{7/6}$ for $M_h=10^9-10^7$ $M_\odot$
(so $Cn^{1/2}<<Bn^{7/18}$).

Evaluation of Equation (6) yields $\gamma -1\sim 0.01-0.5$
for hydrogen densities of $1-10^5$ cm$^{-3}$ for $z=20$ and a $10^8$
$M_\odot$ halo.
Note that a density of $1$ cm$^{-3}$ is achieved for our halos after a
contraction in radius by a factor of a few, much less than the contraction
factor $\lambda^{-1}\sim 20$ after which a disk forms (Mo, Mao \& White 1998).
One finds, for the $10^9$ $M_\odot$ halo at $z=20$ and
for the appropriate (column) density scaling with $r$, e.g.,
$\tau_0\sim r^{-1}$ and $n\sim r^{-2}$, that
$t_{ff}\ge t_{ph}$ and $\gamma\ge 1.1$ for $r\le r_{dyn}=0.02L$. This
implies enclosed masses, $M\sim r$, of about $0.02-0.003M_h$ for the adopted
isothermal profile and halo masses. The adiabatic value $\gamma =4/3$ is
achieved for $r\le r_{dyn}= 0.002L$ and a $10^9$ $M_\odot$ halo at $z=20$
(but see see some corrections to $\gamma$ in the next subsection).

The presence of the 'C term' from Equation (3) does not
mean that conversion of Lyman $\alpha$ photons to the two-photon continuum
is a significant sink. Rather, for the considered halos, trapping of Lyman
$\alpha$ occurs already at densities for which
collisional de-excitation by protons is negligible (despite the large columns
$N_H\sim N_c$ or somewhat smaller) because the thermal electron abundance is
very small\footnote{In fact, at densities below $10^3$ cm$^{-3}$ one has $N_H<<N_c$, but we retain the intuitive form of Equation (6) because the logarithm renders any error insignificant anyway.}. We will come back to the consequences of the
rise in temperature associated with $\gamma >1$ in the next subsection.
Finally, the $7/18$ dependence on density renders our results relatively
insensitive to subtleties in the Lyman $\alpha$ radiative transfer.


\subsection{Dynamic Case: Two-Quantum and HeII Corrections}

We have assumed that the gas remains close to, but not exactly at, $T=10^4$ K
as far as its density profile is concerned.
This is reasonable given the sharpness of the Lyman $\alpha$ cooling
function. A value $\gamma >1$ implies of course that the temperature
rises with increasing density, but Lyman $\alpha$ cooling will
dominate the local thermal balance for temperatures $T<5\times 10^4$ K.
Of course, as the temperature rises, so do the electron and proton abundance
and this favors the two-photon continuum by decreasing $N_c$. From Equation
(6), thermal ionization balance and the results for $N_c(y)$, one finds that
$\gamma$ weakens towards unity above $\sim 2\times 10^4$ K when the electron
abundance exceeds $0.1$. However, this temperature is reached when the density
is $10^{3.5}$ cm$^{-3}$ and two-quantum decay is quickly shutdown during the
collapse as a density of $10^5$ cm$^{-3}$ is exceeded.

In fact, at temperatures above $5\times 10^4$ K HeII line cooling dominates,
and the latter also suffers from photon trapping when hydrogen columns exceed
$10^{24}$
cm$^{-2}$. That is, the HeII Lyman $\alpha$ line at 304\AA\ is similarly
opaque (Neufeld 1990), barring the appropriate changes in Einstein A
coefficient, elemental abundance etc., as its HI counterpart. One has that
$\tau_{He} =5.2\times 10^{-14}N(He^+)T_4^{-1/2}$.
Hence, photon trapping will continue, for
the massive halos that we consider, into the HeII regime at large columns.
Also, the much larger HeII Lyman $\alpha$ Einstein A coefficient of
$\sim 10^{10}$ s$^{-1}$ boosts the required value of $y$ for a given $N_c$
by two orders of magnitude. The HeII two-photon channel
is shut down since densities exceed $10^{5.5}$
cm$^{-3}$ around the HeII cooling peak ($A_{2s1s}\sim 8.2Z^6$ s$^{-1}$).
In any event, most HeII two-quantum decay photons are absorbed by the (HI Lyman
$\alpha$ trapping) neutral hydrogen that surrounds the halo core.

As a result of all this $\gamma$ remains well above unity and the
system evolves adiabatically for densities above $\sim 10^5$ cm$^{-3}$.
Equation (6) provides a good fit to the atomic physics of HI and HeII between
$n=1$ and $n=10^7$ cm$^{-3}$ if corrected for HI two-photon decay and
the change in line opctical depth as HI cooling is superseded by HeII cooling.
One finds that
$$\gamma -1\approx -{{{{1}\over{2}}+{{7}\over{18}}B'(n)n^{7/18}}\over{log(Cn^{1/2})+B'(n)n^{7/18}}},\eqno(7)$$
where $B'=0$ for $n=10^3(M_h/10^9M_\odot)^{-1/3}-10^5$ cm$^{-3}$
following Equation (3), and where $B'\approx 0.36B$ if $n\ge n_c=10^{5.5}$
cm$^{-3}$ and $B'=B$ otherwise.
Note here that $t_{ph}>t_{ff}$ always holds and that $B'n^{7/18}>>1$ for
densities larger than $n_c$, and for all halos, i.e., the HI to HeII switch
has a modest impact because the optical depth enters into $t_{ph}$ with a $1/3$
power.

\section{Discussion and Future Work}

The Jeans mass
for a $0.1$ cm$^{-3}$ halo is about $M_J\sim 3\times 10^7$ $M_\odot$
at $10^4$ K. Over the Lyman $\alpha$ cooling regime, $M_J$
decreases only by a factor of about $8$. That is, a $0.02-0.003M_h$ core
will likely not experience significant fragmentation during
gravitational collapse up to densities of $\sim n(r_c)\approx 10^5$
cm$^{-3}$ at $z=10-20$, after which fragmentation is halted
adiabatically. That is, the system cannot cool above a few times $10^5$ K
either because photons produced by, e.g., bremsstrahlung cannot escape
since the bulk of these cooling
photons are at energies above a few times $10^{15}$ Hz and are
re-proprocessed into Lyman alpha and trapped in the surrounding
neutral exterior of the collapsing cloud.
Also, the rise in $\gamma$ is moderate enough to justify our
use of an isothermal density profile.

Although our results are order of magnitude estimates, they connect quite
well with the detailed numerical simulations of gravitational collapse
by Jappsen et
al.\ (2005) and Klessen, Spaans \& Jappsen (2005). The former authors find
that a switch to a $\gamma >1$ region in density space for a collapsing gas
sets a characteristic mass scale for fragmentation through the Jeans mass at
the ambient density and temperature. Hence, a value $\gamma > 1$
appears to be a robust indicator of the lack of fragmentation.
As such, the picture that emerges from detailed hydrodynamical simulations
and the shape of the EOS is at least consistent.

Finally, the frequency shift that a Lyman $\alpha$
photon experiences before escape scales approximately as $\nu_{\rm
shift}\sim T^{1/6}$ (Dijkstra et al.\ 2006). Hence, fluctuations in
temperature do not strongly influence our results in this respect either.

Thus, allowing for some expected inefficiency, we infer that of order 0.1\%
of the baryon mass forms a pregalactic black hole of mass
$M_{BH}\sim 10^4-10^6$ $M_\odot$. Note here that Bromm \& Loeb (2003) find a
similar inhibition of fragmentation from detailed hydrodynamic simulations
that assume a rougly isothermal ($\gamma\sim 1$) collapse. They do not
include the trapping effects discussed here. So unless there are numerical
resolution effects that play a role, a value of $\gamma =1$ may already be
sufficient to halt fragmentation.

In any case, these so-called intermediate mass black holes
(IMBH) are plausible seeds for generating the supermassive,
$\sim 0.001 M_{spheroid}$, black holes found in galaxy cores today
(c.f.\ H\"aring \& Rix 2004) by gas accretion (Islam, Taylor \& Silk 2003).
The inferred presence of pregalactic IMBH and their associated accretion
luminosity has been a source of intense speculation
with regard to a mechanism for the reionization of the universe
(e.g., Madau et al.\ 2004; Ricotti \& Ostriker 2004;
Venkatesan, Giroux \& Shull 2001).
Our results place these speculations on a sounder footing. 
Moreover, isolated IMBH should 
exist in galactic halos at a similar mass fraction
according to simple models for generating the Magorrian correlation 
between central black hole mass and spheroid velocity dispersion
(Islam, Taylor \& Silk 2004; Volonteri, Haardt \& Madau 2003)
and possibly be detectable as gamma-ray sources (Zhao \& Silk 2005).

Still, there are a number of other issues that should be addressed in the
future.

1a) The stiffening of the polytropic equation of state found in this work
depends crucially on the absence of any H$_2$ molecules. For temperatures
above 3,000 K this seems plausible because collisional dissociation and
charge exchange with H$^+$ limit the abundance of H$_2$,
while Lyman $\alpha$ trapping keeps the temperature above $10^4$ K.
Still, H$_2$ may also form directly in massive halos with virial
temperatures above $10^4$ K (Oh \& Haiman 2002) and reach a universal
abundance of $\sim 10^{-3}$. H$_2$ formation in these cases is a consequence of
a freeze-out of the H$_2$ abundance, in the presence of a large free electron
fraction, as gas cools from above $10^4$ K on a time scale that is shorter than
the H$_2$ dissociation time. However, in the dense and massive halos that
we consider Lyman $\alpha$ trapping,
already at modest densities of $1$ cm$^{-3}$, causes the cooling time to
increase exponentially, from a level of a few $\times 10^6$ yr at 8,000 K,
and to remain larger than the H$_2$ dissociation time. That is,
the gas lingers at $10^4$ K, unable to reach 8,000 K or less, and stays
at those temperatures because H$_2$ formation is suppressed (H$_2$ is easily
destroyed by H$^+$) at the ambient temperatures (Oh \& Haiman, reaction 17 on
their page 15). As a consequence the H$_2$ abundance remains
at a very low level around $10^4$ K, see Figure 4 of Oh \& Haiman (2002),
and does not contribute to the cooling.
Furthermore, Figure 7 of Bromm \& Loeb (2003) shows that even a halo with
$T_{vir}\sim 10^4$ K (baryonic mass of $\sim 10^7$ $M_\odot$) first heats a
large part of the cold (30-100 K) infalling gas to temperatures
of $3,000-10,000$ K for densities of $\sim 1$ cm$^{-3}$, when the formation of,
and cooling by, H$_2$ is incorporated.
This is the relevant, minimum temperature range because there is a trough
between the Lyman $\alpha$ and H$_2$ cooling curves here (Oh \& Haiman 2002).
In this, it is important to realize that our halos are quite massive, baryonic
masses of $10^7-10^9$ $M_\odot$, and dense ($z>10$). These values
favor Lyman $\alpha$ trapping.

1b) Still,
the presence of a UV background from popIII stars, that suppresses
the abundance of H$_2$ molecules (Bromm \& Loeb 2003), would certainly be
welcome. The critical density for collisional H$_2$ dissociation to dominate
is about 300 cm$^{-3}$, if a UV background as in, e.g., Bromm \& Loeb (2003)
is present with which H$_2$ collisional dissociation has to compete.
We do reach this regime early in the collapse so that self-shielding effects
would not limit the benefits of H$_2$ photo-dissociation much (see Bromm \&
Loeb 2003).
Of course, in the absence of a background radiation field, all the
time scales (formation, dissociation, etc.) in the system scale as 1/density
and thus their ratios are independent of density and the discussion of 1a
applies.

1c) The formation of the black hole will introduce a quasar
whose power-law spectral energy distribution can boost the formation of H$_2$
through the H$^-$ route (Haiman, Abel \& Rees 2000).
Hence, for redshifts below $\sim 300$, where H$^-$ is
no longer destroyed by the CMB, black hole formation as discussed here will
facilitate the formation of H$_2$ and impact the EOS of the gas
surrounding the black hole (Scalo \& Biswas 2002). The mode that we describe
here would then be inhibited unless an H$_2$ dissociating UV background as
in Bromm \& Loeb (2003) or Oh \& Haiman (2002) is present.
Still, the large columns that we consider
would shield at least part of the gas from X-ray feedback (see 2b below).

2a) Trace amounts of dust as little as $10^{-4}$ of Solar are sufficient to
absorb all Lyman $\alpha$ photons in a homogeneous halo for the columns
considered in this work (Neufeld 1990). Hence, the
stiffening of the EOS that we have found
disappears once the first metals have been produced because dust
emission is optically thin. Also, metals are efficient coolants and,
if present, would take over the cooling for radial contraction factors larger
than $60$ (Haiman \& Spaans 1999).
Inhomogeneity suppresses dust absorption, but
facilitates the escape of Lyman $\alpha$ photons by boosting $\epsilon$
(Haiman \& Spaans 1999). In any case, the formation of these massive
black holes is stopped once the ambient
metallicity increases due to star formation. Hence, the fraction of massive
primordial galaxies that harbor these black holes is dictated by the
fraction of metal-free gas at $z=10-20$. Given that popIII star formation
is co-eval with this epoch, the metal-free gas fraction is uncertain and
clumpy (Scannapieco, Schneider \& Ferrara 2003). Hence, the overall
contribution of this mode of black hole formation is somewhat undetermined
and likely to lie anywhere between 5 and 50 percent, depending on the
proximity of other galaxies. If efficient, this mode may violate the
three-year WMAP constraint on the electron scattering optical depth of
$\tau_e\approx 0.09$ (Spergel et al.\ 2006), because of the large X-ray output
expected for these massive and late black holes.

2b) Fortunately, following Ricotti et al.\
(2005) it is possible to constrain the contribution to $\tau_e$. The latter
authors find that $\tau_e$
scales as $1/log(N_H)$ and levels off to $\tau_e\sim 0.1$ for columns in
excess of $10^{22}$ cm$^{-2}$. Given that all our black holes form as
the end product of a central collapse from a massive halo, the surrounding
gas has column densities between $\sim 10^{22}$ cm$^{-2}$ (from the initial
halo masses and redshifts) and $\sim 10^{26}$ cm$^{-2}$ (from the $\gamma =4/3$
adiabatic points). We have used the models of Meijerink \& Spaans (2005) to
confirm that these columns are sufficient to re-process X-rays between 1
and 30 keV for metallicities in the surrounding gas between 0 and $10^{-2}$.
Hence, a value of $\tau_e\sim 0.1$ is appropriate for our black holes,
even if they would be the dominant mode of black hole formation.

3) We assume zero angular momentum for the gas, but the radiative transfer
is not sensitive to the associated velocity field. Of course, our
arguments are purely thermodynamic in nature and do not solve
the angular momentum problem if the initial cloud is rotating.

4) We would expect that dwarf spheroidals should have central black
holes in the range of $10^4-10^6$ $M_\odot$, whereas irregulars, and
in particular late-forming dwarfs, should not have such central IMBH.
In order to substantiate this, more detailed hydrodynamical simulations
that include dynamical photon trapping, or its EOS parameterization, should
be performed.

\acknowledgments

JS gratefully acknowledges the hospitality of the Kapteyn Astronomical
Institute as a Blaauw Visiting Professor, where this work was initiated.
We are thankful to an anonymous referee for his insightful comments.
We thank Avi Loeb and Volker Bromm for discussions on black hole formation
and Leonid Chuzhoy for discussions on Lyman $\alpha$ trapping.


\begin{thebibliography}{}

\bibitem[]{}Abel, T., Bryan, G.L., \& Norman, M.L., 2000, ApJ, 540, 39

\bibitem[]{}Abel, T., Bryan, G.L., \& Norman, M.L., 2002, Science, 295, 93

\bibitem[]{}Bromm, V. \& Loeb, 2003, ApJ, 596, 34

\bibitem[]{}Bromm, V., Coppi, P.S., \& Larson, R.B., 2002, ApJ, 564, 23

\bibitem[]{}Dijkstra, M., Haiman, Z., \& Spaans, M., 2006, astro-ph/0510407

\bibitem[]{}Haiman, Z. \& Spaans, 1999, ApJ, 518, 138

\bibitem[]{}Haiman, Z., Spaans, M., \& Quataert, E., 2000, ApJ, 537, L5

\bibitem[]{}Haiman, Z., Abel, T. \& Rees, M.J., 2000, ApJ, 534, 11

\bibitem[]{}Haiman, Z., Rees, M.J. \& Loeb, A., 1997, ApJ, 476, 458

\bibitem[]{}H\"aring. N. \& Rix, H.-W., 2004, ApJ, 604, L89

\bibitem[]{}Islam, R.,  Taylor, J., \& Silk, J.  2004, MNRAS, 354, 427I

\bibitem[]{}Islam, R., Taylor, J., \& Silk, J.  2003, MNRAS, 340, 647I

\bibitem[]{}Jappsen, A.-K., Klessen, R.S., Larson, R.B., Li, Y. \& Mac Low, M.-M., 2005, A\&A, 435, 611

\bibitem[]{}Klessen, R.S., Spaans, M. \& Jappsen, A.-K., 2005, IAU Symp. 227, 337

\bibitem[]{}Koushiappas, S.M., Bullock, J., \& Dekel, A., 2004, MNRAS, 354, 292

\bibitem[]{}Li, Y., Mac Low, M.-M., \& Klessen, R.S., 2003, ApJ, 592, 975

\bibitem[]{}Madau, P., Rees, M., Volonteri, M., Haardt, F., \& Oh, S. 2004, ApJ, 604, 484.

\bibitem[]{}Meijerink, R. \& Spaans, M., 2005, A\&A, 436, 397

\bibitem[]{}Mo, H.J., Mao, S. \& White, S.D.M., 1998, MNRAS, 295, 319

\bibitem[]{}Neufeld, D.A., 1990, ApJ, 350, 216

\bibitem[]{}Oh, S.P. \& Haiman, Z., 2002, ApJ, 569, 558

\bibitem[]{}Portegies Zwart, S.F., Baumgardt, H., Hut, P., Makino, J., \& McMillan, S.L.W., 2004, Nature, Vol 428, Issue 6984, p.\ 724

\bibitem[]{}Rees, M.J. \& Ostriker, J.P., 1977, MNRAS, 179, 541

\bibitem[]{}Ricotti, M. \& Ostriker, J.P., 2004, MNRAS, 352, 547

\bibitem[]{}Ricotti, M., Ostriker, J.P. \& Gnedin, N.Y., 2005, MNRAS, 357, 207

\bibitem[]{}Scalo, J. \& Biswas, A., 2002, MNRAS, 332, 769

\bibitem[]{}Scannapieco, E., Schneider, R. \& Ferrara, A., 2003, ApJ, 589, 35

\bibitem[]{}Silk, J., 1977, ApJ, 211, 638

\bibitem[]{}Silk, J., 2005, MNRAS, 364, 1337

\bibitem[]{}Spaans, M. \& Silk, J., 2000, ApJ, 538, 115

\bibitem[]{}Spaans, M. \& Silk, J., 2005, ApJ, 626, 644

\bibitem[]{}Spergel, D.N., Bean, R., et al., 2006,
http://lambda.gsfc.nasa.gov/product/
map/dr2/pub$_{}$papers/threeyear/parameters/wmap$_{}$3yr$_{}$param.pdf

\bibitem[]{}Spitzer, Ly.\ Jr.\, 1978, in 'Physical Processes in the Interstellar Medium' (Wiley)

\bibitem[]{}Venkatesan, A., Giroux, M.L. \& Shull, J.M., 2001, ApJ, 563, 1

\bibitem[]{}Volonteri, M.,  Haardt, F., \& Madau, P., 2003, ApJ, 582, 559

\bibitem[]{}Zhao, H. \& Silk, J., 2005, PRL, 95, 011301

\end{thebibliography}
\end{document}